\documentclass[11pt]{article}
\usepackage{graphics,epsfig}
\usepackage[latin1]{inputenc}
\usepackage[english,activeacute]{babel}
\usepackage[]{graphicx}
\usepackage{latexsym}
\usepackage{amsmath, amsthm, amsfonts, amssymb}
\usepackage{verbatim}

\begin{document}

\newtheorem{theo}{Theorem}[section]
\newtheorem{definition}[theo]{Definition}
\newtheorem{lem}[theo]{Lemma}
\newtheorem{prop}[theo]{Proposition}
\newtheorem{coro}[theo]{Corollary}
\newtheorem{exam}[theo]{Example}
\newtheorem{rema}[theo]{Remark}
\newtheorem{example}[theo]{Example}
\newtheorem{principle}[theo]{Principle}
\newcommand{\ninv}{\mathord{\sim}}
\newtheorem{axiom}[theo]{Axiom}

\title{Logical Structures Underlying Quantum Computing}

\author{Federico Holik $^{1}$, Giuseppe Sergioli $^{2}$, Hector
Freytes $^{2}$ and Angel Plastino $^{1}$}

\maketitle

\begin{center}

\begin{small}
1- Instituto de Fisica (IFLP-CCT-CONICET), Universidad Nacional de
La Plata, C.C. 727, 1900 La Plata, Argentina; holik@fisica.unlp.edu.ar (F.H.); angeloplastino@gmail.com (A.P.)\\
2- Universit\`{a} di Cagliari, Via Is Mirrionis 1, 09123 Cagliari,
Italy; giuseppe.sergioli@gmail.com (G.S.); hfreytes@gmail.com
(H.F.).
\end{small}
\end{center}

\vspace{1cm}

\begin{abstract}
\noindent In this work we advance a generalization of quantum
computational logics capable of dealing with some important examples
of quantum algorithms. We outline an algebraic axiomatization of
these structures.
\end{abstract}
\bigskip
\noindent

\begin{small}
\centerline{\em Key words: quantum computing; non-Kolmogorovian
probability; quantum computational gates}
\end{small}

\section{Introduction}

Quantum computers are one of the main technological goals of our
time \cite{NielsenandChuang} and many efforts are focused on their
development. While some examples of quantum computers were actually
built~\cite{Gambetta-Review}, they are still not strong enough to
overcome their classical competitors: decoherence poses a threat to
the problem of scaling up the number of qubits
\cite{NielsenandChuang}. While difficult to implement, the results
are promising, and a lot of interest revolves around the study of
quantum algorithms both, from~a theoretical standpoint and an
empirical one as well. In this paper, we focus on the following
problem: the characterization of the logical and algebraic
structures underlying quantum algorithms. The~characterization of
these structures is of fundamental importance for understanding the
peculiarities of quantum computation.

To describe quantum computation from a logical and algebraic point
of view, many formalisms were developed (see for example
\cite{DF,DallaChiara-Giuntini-Sergioli-2013,DallaChiara-IJQI,FS,ManyValuedApproach,EntropyDallachiara,FirstOrder}).
We refer to the logics studied in these works as \emph{{quantum
computational logics}} (QCL). The QCL approach is based on the
characterization of the action of quantum gates using probabilistic
truth values (see also \cite{FS}).

In this work, we present a generalization of the QCL approach in
which the truth values are extended to include arbitrary readings of
the quantum register. This move allows us to represent quantum
algorithms of practical interest in a natural way. We also
generalize the QCL approach to a huge family of propositional
systems, based on orthomodular lattices. Putting the emphasis in the
propositional structure of the readouts of the quantum register
allows for a better comparison with the classical case. Our
generalization reveals that \emph{quantum computing can be
considered as a non-Kolmogorovian version of classical
non-deterministic computing}, continuing the line of research
proposed in \cite{HolikEntropy}. From this perspective, the
orthomodular lattice of projection operators of the Hilbert space is
the essential algebraic structure. This lattice was termed
\emph{quantum logic} after the pioneer work of Birkhoff and von
Neumann \cite{BvN} (for a more recent approach, c.f.
\cite{Holik-AOP-2014,Holik-Massri-JMP,Holik-Ciancaglini,Holik-Zuberman-2013}).
In the quantum case, the non-Kolmogorovian character of the
probabilistic calculus involved, comes directly from the fact that
there are complementary contexts in which measurements can be
performed. We think that this is a very important point, because it
opens the door to investigating the role of contextuality in QC in a
more explicit and natural way. This is in harmony with recent
studies that suggest that the essential resource allowing for the
advantages of quantum computing is contextuality
\cite{Contextuality-Anders,Contextuality-Delfosse,Contextuality-Howard,Contextuality-Raussendorf,Contextuality-RaussendorfPRA},
a physical phenomenon strongly related to the non-distributive
character of the lattice of quantum propositions
\cite{Doring-KSVNA}. See also \cite{Bub-QC}, for a discussion of the
fundamental role played by the projective geometry of the Hilbert
space in quantum algorithms. Another advantage of our generalization
is that of knowing a canonical form of finding quantum versions of
classical formalisms and algorithms.

Finally, our approach shows how to conceive alternative forms of
non-classical computing. Indeed, the general scheme presented in
this manuscript could be used to compare QC with other alternative
theories. A comparison of this kind, could help us to understand
better the nature of quantum computing. Notice also that our
formalism can be of guide for defining algorithms in physical
frameworks that go beyond standard quantum mechanics. Indeed, the
rigorous formulation of quantum field theory and quantum statistical
mechanics require factor von Neumann algebras that go beyond the
standard Type I case \cite{Redei-Summers2006}, reflecting the fact
that these are (from a technical standpoint) non-equivalent
probabilistic theories. Our approach can be of use in the study of
quantum computing in the limit of systems with infinitely many
degrees of freedom.

The paper is organized as follows. In Section
\ref{s:ClassicalComputing} we review some general aspects of
classical computing to motivate our further developments
and present the essential aspects of the different forms of
computing in a schematic way. We present our generalization of
quantum computational logic in Section
\ref{s:QuantumComputationalLogics}. Next, in Section
\ref{s:QuantumAlgorithms} we review how concrete algorithms of
interest in applications fall into our theoretical scheme. In
Section \ref{s:Axiomatization} we provide the outline of an
axiomatic framework for these logics, which contain classical and
quantum computing as particular cases. We~also discuss how these
algebraic structures can be expressed in math-fashion. Finally, some~conclusions are drawn in Section~\ref{s:Conclusions}.

\section{Classical Computing}\label{s:ClassicalComputing}

A classical computing machine is based on bits. The bit is the
elementary unit for measuring information. A physical bit (a system
with only two relevant states) can take two values $0$ and $1$. In~general, information will be stored and processed in classical
computers by appealing to physical bits and operations on them. In
this Section we start by reviewing the notions of deterministic and
non-deterministic classical computation. Next, we describe the
essential aspects of some particular models of quantum computers.

\subsection{Deterministic Classical Computing}

Any function $F:\{0,1\}^{N}\longrightarrow \{0,1\}^{M}$ can be
expressed in terms of Boolean functions (see for example
\cite{Kitaev-Book}, Chapter {2}). Define a Boolean function
as a function $F:\{0,1\}^{N}\longrightarrow \{0,1\}$. A
\emph{Boolean circuit} can be represented as a composition of
elementary Boolean functions. In general, a basis $\mathcal{A}$ of
elementary functions is taken as a generating set for all other
Boolean functions. One of the most important examples is the choice
$\mathcal{A}_{1}=\{\vee,\wedge,\neg\}$, where

\begin{definition}\label{d:Disjunction}
$\vee:\{0,1\}^{2}\longrightarrow\{0,1\}$ with
$$\vee(0,0)=0$$
$$\vee(0,1)=1$$
$$\vee(1,0)=1$$
$$\vee(1,1)=1$$
$\wedge:\{0,1\}^{2}\longrightarrow\{0,1\}$ with
$$\wedge(0,0)=0$$
$$\wedge(0,1)=0$$
$$\wedge(1,0)=0$$
$$\wedge(1,1)=1$$
$\neg:\{0,1\}\longrightarrow\{0,1\}$ with
$$\neg(0)=1$$
$$\neg(1)=0$$
\end{definition}

\noindent Notice that the elements of $\mathcal{A}_{1}$ have
different $n$-arity. Any Boolean function
$F:\{0,1\}^{N}\longrightarrow \{0,1\}$ can be written as a
composition of the elementary functions belonging to
$\mathcal{A}_{1}$. Thus, the computation of $F$ can be effected via
a physical circuit made up physical representations of elementary
Boolean functions (elementary classical gates). In other words, the
hardware that allows us to compute the function $F$ can be made up
by electronic components representing the elementary functions
$\vee$, $\wedge$ and $\neg$, combined in a suitable way. The
generalization of Boolean functions to functions
$F:\{0,1\}^{N}\longrightarrow \{0,1\}^{M}$ (with $M\geq 2$) is
straightforward (we refer the reader to \cite{Kitaev-Book} for
details).

The problem of finding a suitable notion of equivalence between
circuits is of major importance. In general, there is more than one
way to represent a function $F$ as a composition of elementary
Boolean functions (and thus, very different hardware devices may
compute exactly the same function). Given two functions
$F,G:\{0,1\}^{N}\longrightarrow \{0,1\}^{M}$, how can we determine
whether they are equal or not? Notice that similarly, we can ask if
two circuits, implementing functions $F$ and $G$, are essentially
the same or not. This is equivalent to testing the proposition
$F=G$. In the classical case, this can be solved in a trivial way by
simply considering all possible inputs $x\in\{0,1\}^{N}$ and
checking whether the outputs $F(x)$ and $G(x)$ are equal or not.
Equivalently, if we do not know the functions, but we have the
hardware representing them, we can compare them in a similar way, by
running the two circuits and comparing the outputs. Notice that this
procedure is equivalent to computing truth values of the elementary
functions: if we know the truth tables of all Boolean generators, we
will be able to compute the outputs of any Boolean function.

Please note that there is still more structure involved in this comparison
between $F$ and $G$. $\{0,1\}^{N}$ is a set, and the set of all its
subsets $\mathcal{P}(\{0,1\}^{N})$ is a Boolean algebra (with the
conjunction $\wedge$ taken as set intersection, the disjunction
$\vee$ as set union and the negation $\neg$ as the set theoretical
complement). This is of major importance for understanding the
extension to classical probabilistic computing and quantum
computing. Let us discuss this with more detail. A rational agent
whose function is to manage the readout of the register can only
deal with (and communicate) truth values of the propositional
structure defined by the power set $\mathcal{P}(\{0,1\}^{N})$. In
this sense, the logic associated with a classical computer is
represented by a Boolean algebra.

\subsection{Non-Deterministic Classical Computing}

The introduction of probabilistic steps in a computation proves to
be useful for solving many particular problems \cite{Kitaev-Book}.
Indeed, the exact solution of some problems displays high
computational complexity, but this complexity can be lowered if we
allow for a low rate of errors in the computation.

However, if the steps of the computation are produced in a probabilistic
way, the output of an input $x\in\{0,1\}^{M}$ must be described by a
probabilistic function $F_{x}:\{0,1\}^{N}\longrightarrow[0,1]$
satisfying
\begin{eqnarray}\label{e:NDCNormalized}
&\forall\,\,x\in\{0,1\}^{M}\,\,&\nonumber\\
&\sum_{y\in\{0,1\}^{N}}F_{x}(y)=1&
\end{eqnarray}

\noindent Here arises an important observation. In the previous
Section, we showed that the propositional structure associated with a
rational agent dealing with the register was exactly the Boolean
algebra $\mathcal{P}(\{0,1\}^{N})$. However, Equation (\ref{e:NDCNormalized}) defines in a canonical way (for each $x\in\{0,1\}^{M}$) a
Kolmogorovian probability distribution in $\mathcal{P}(\{0,1\}^{N})$
as follows:

$$\mu_{F_{x}}:\mathcal{P}(\{0,1\}^{N})\rightarrow[0,1]$$

\noindent such that:

\begin{enumerate}
\item $\mu_{F_{x}}(\emptyset)=0$
\item $\mu_{F_{x}}(A^{c})=1-\mu_{F_{x}}(A)$
\item For any disjoint $X,Y\in\mathcal{P}(\{0,1\}^{N})$ we have
$$\mu_{F_{x}}(X\bigcup Y)=\mu_{F_{x}}(X)+\mu_{F_{x}}(Y)$$
\end{enumerate}
\noindent Thus, the classical propositional structure of the
register implies that there will be a classical probabilistic
distribution for the propositions communicated by a rational agent
dealing with the readouts.

\subsection{Quantum Computing in a Schematic Way}\label{s:QuantumComputers}

Before we continue, it is useful to give some definitions. States of
a quantum system can be represented by so-called \emph{density operators},
which are positive and trace class self-adjoint operators of trace
one, acting on a separable Hilbert space $\mathcal{H}$. We call
$\mathcal{C}(\mathcal{H})$ to the set of all density operators. A~density operator $\rho$ is said to be \emph{pure} if $\rho^{2}=\rho$
and \emph{mixed} otherwise. An operator $U$ acting on a separable
Hilbert space is said to be \emph{unitary} iff
$UU^{\dag}=U^{\dag}U={\mathbf{1}}$ (here, $U^{\dag}$ represents the 
adjoint of the operator $U$). Quantum gates will be represented by
unitary operators. As is well known, if $U$ is unitary, the map
$\mathcal{E}(\rho)=U\rho U^{\dagger}$ maps density operators into
density operators.

Let us try to summarize how a quantum computer works. Suppose that
our quantum hardware has $N$ qubits. In this case, the Hilbert space
is $\mathcal{H}=\bigotimes^{N}\mathbb{C}^{2}$. A basic general
description of all the central operations needed to perform a
quantum algorithm can be given as follows.

\begin{itemize}
\item {Step 1}. Chose an initial state represented by a density operator
$\rho$ for the qubits (notice that $\rho$ can be taken to be mixed
\cite{AKN}). This state could be just the density operator
associated with the vector
$|0\rangle\otimes|0\rangle\otimes\cdots\otimes|0\rangle$ or any
other desired state. In most situations of practical interest,
the~computational basis plays a key role in defining the inputs and
the outputs of the algorithm. Let us denote the computational basis
by $\mathbf{B}_{0}$.
\item {Step 2}. Apply a collection of gates represented by
unitary operators $\{U_{i}\}_{i=1,...,n}$ to reach a
desired final state $\sigma=U_{n}\cdots U_{2}U_{1}\rho
U_{1}^{\dagger}U_{2}^{\dagger}\cdots U_{n}^{\dagger}$.
\item {Step 3}. Perform a measurement on the system when state
$\sigma$ is reached, check the result obtained, and depending on the
result, stop the process, or continue following the pertinent
protocol if necessary (which will involve a similar, or, in some
cases, the same- process). Notice that the measurement process
involves the choice of a measurement basis. This process has
associated a probability (dictated by Born's rule). The probability
of success for an algorithm is related to the probability of
occurrence of a certain outcome (or collection of them). This
outcome should be of interest for the successful computation
involved in the protocol that one is following. Notice that a
subspace containing the desired results always exists. The pertinent
probability of success is related to the probability for the event
of interest to happen.
\end{itemize}

It is illustrative to compare now the possible readouts of a
rational agent $R$ dealing with a quantum computer. Like in the
classical probabilistic case, the possible observations are not
restricted to the computational basis $\mathbf{B}_{0}$. Notice that
$\mathbf{B}_{0}$ defines a Boolean algebra of operators in a
canonical way as follows. First, consider the set $\mathbf{P}_{0}$
of all possible orthogonal projection operators corresponding to the
elements of $\mathbf{B}_{0}$. The commutant of a set $C$ of bounded
operators acting on a Hilbert space is defined as
$C':=\{x\in\mathcal{B}(\mathcal{H})\,\,|\,\,[x,y]=0\,\,\forall\,\,y\in
C\}$. The double commutant of $C$ is defined as $(C')'$. Now, take
the double commutant of $\mathbf{P}_{0}$ to define the set
$\mathcal{P}_{\mathbf{B}_{0}}:=(\mathbf{P}_{0})''$. It is easy to
check that $\mathcal{P}_{\mathbf{B}_{0}}:=(\mathbf{P}_{0})''$ is a
Boolean algebra. The projection operators in
$\mathcal{P}_{\mathbf{B}_{0}}$ form the propositional structure
associated with the computational basis. However, as in the
classical case, the measurements of the rational agent will not be
generally restricted, to this set of propositions. Indeed, by
applying rotations on the output state, $R$~can measure other
properties associated with the final quantum state. In case that,
due to restrictions in the hardware, the readouts are only performed
on the computational basis, a measurement in a different basis can
be implemented by a rotation $U$ applied on the quantum state (using
the equivalence $\mbox{Tr}(\rho U^{\dagger}PU)=\mbox{Tr}(U\rho
U^{\dagger}P)$). As it is well known, the set of possible
propositions to be tested is formed by the set
$\mathcal{P}(\mathcal{H})$ of all orthogonal projection operators
acting in $\mathcal{H}$. The collection $\mathcal{P}(\mathcal{H})$
is a modular non-distributive lattice for finite dimensional Hilbert
spaces and an orthomodular one for infinite dimensional ones. The
final quantum state---represented by the density operator
$\sigma$---defines a quantum probability distribution
$$P_{\sigma}:\mathcal{P}(\mathcal{H})\longrightarrow [0,1]$$

\noindent given by $P_{\sigma}(P)=\mbox{Tr}(\sigma P)$ for all
$P\in\mathcal{P}(\mathcal{H})$. It possesses the following
properties:

\begin{enumerate}
\item [1] $P_{\sigma}(\mathbf{0})=0$ ($\textbf{0}$ is the null operator).
\item [2] $P_{\sigma}(P^{\bot})=\mathbf{1}-P_{\sigma}(P)$, for all
$P\in\mathcal{P}(\mathcal{H})$.
\item [3] For any family $(P_{j})_{j\in J}\subset\mathcal{P}(\mathcal{H})$ consisting
of orthogonal projections for which $P_{j_{1}}P_{j_{2}}=\mathbf{0}$
when $j_{1}\neq j_{2}$, the following equality holds:
$$P_{\sigma}(\bigvee P_{j})=\sum_{j\in J}
P_{\sigma}(P_{j}).$$
\end{enumerate}
\noindent Observe that, since the Hilbert space $\mathcal{H}$ is
supposed to be separable, the family  may be taken to be finite or
countable. Notice also that the equality $P \vee Q + P \wedge Q = P
+ Q$ is true provided $PQ = QP = P \wedge Q$. Here the operations $P
\vee Q$ and $P \wedge Q$ are appropriately defined within the
$W^{*}$-algebra (or von Neumann algebra) under discussion. In case
that this $W^{*}$-algebra coincides with $\mathcal{B}(\mathcal{H})$,
i.e., the algebra of all bounded linear operators on the Hilbert
space $\mathcal{H}$, then the operator $P\wedge Q$ is the orthogonal
projection on the subspace $P\mathcal{H}\cap Q\mathcal{H}$, and $P
\vee Q$ is the orthogonal projection on the closure of the subspace
$P\mathcal{H} + Q\mathcal{H}$. Of course, similar equalities hold for
families of commuting orthogonal projections. We~always have
$P^{\bot}=\mathbf{1}-P$

While the properties of a quantum probability distribution
may resemble the properties of a Kolmogorovian one, there is a
radical difference. The probability distribution associated with a
classical probabilistic algorithm is defined in a Boolean algebra,
but the probability distribution associated with a quantum one is
defined in an orthomodular lattice \cite{KAL}. The differences
between these probability theories has been extensively studied  in
the literature (see for example \cite{Gudder-StatisticalMethods,Holik-AOP-2014,Redei-Summers2006}). For more
discussion on this subject see
\cite{Khrennikov-2010,Khrennikov-2016,Aerts-2013}, where the authors
propose applications of non-Kolmogorovian probability theory outside
of the quantum domain.

\section{Logics Associated with Quantum
Algorithms}\label{s:QuantumComputationalLogics}

In this Section we introduce our proposal for generalizing quantum
computational logics. We~start with the characterization of
compositional gates and end up with a generalization of
probabilistic truth values.

\subsection{The Algebraic Structure of Quantum Logical
Gates}\label{s:AlgebrasOfGates}

In a real quantum computer, a general quantum gate is constructed by
adequately composing elementary ones. In theoretical quantum
computing, certain sets of universal gates are used. As~examples, we
can take the quantum Toffoli and the Hadamard gates
\cite{Aharonov1}; another example is given by CNOT, Hadamard, and
controlled phase gates. The important point to remark is the fact
that we use, as a starting point, a set of generating gates
$G=\{G_{1},....,G_{N}\}$ (with finite elements). In the above
examples we have: $G_{1}=\{T,H\}$ and $G_{2}=\{CNOT,H,R_{\phi}\}$.
The native gates of the hardware presented in \cite{Gambetta-Review}
are given by $G_{3}=\{XX,R_{\phi}\}$. In this paper, we concentrate
on $G_{1}$, but a similar analysis could be carried out for $G_{2}$
and $G_{3}$.

Given a generating set of gates $G$, all physically implementable
gates will be given by successive applications of the elements of
$G$. This means that any actual gate implemented in the quantum
computer will have the form: $U=U_{n}\ldots U_{1}$, where $U_{i}\in
G$. Let us call to these compositions of gates (in analogy with the
classical case) \emph{quantum polynomial gates}. Denote by $P(G)$
the set of all possible quantum polynomial gates. If $G$ is formed
by a universal set of gates, then $P(G)$ will be dense in
$\mathcal{U}_{N}(\mathcal{H})$ (the set of unitary operators acting
on $\mathcal{H}$). From a theoretical perspective, this is all we
need to implement any desired quantum algorithm.

Any component of a quantum algorithm will be described as a
succession of polynomials in $P(G)$, and eventually, a reading
(measurement) described by a projection onto a subspace or a family
of them, with their associated probabilities computed using the Born
rule. Notice that $P(G)$ and its topological closure are the quantum
versions of the logic of a classical computer. In the classical
version, we have a functional calculus based on Boolean functions.
The Boolean character is related to the fact that these functions
commute. In the quantum version, we have a non-commutative matrix
calculus instead.

An important problem in the logical approach to classical
computation is given by testing functional equations. This problem
can be solved by appealing to truth tables of elementary Boolean
functions, such as $\vee$, $\wedge$ and $\neg$, as explained in
Section \ref{s:ClassicalComputing}. However, in the quantum version, the
existence of non-compatible context and indeterminate processes
leads us to a different notion of truth, based on probabilistic and
contextual truth values. In the following Subsection, we explain how
this works.

\subsection{Probabilistic Truth Values}\label{s:RingOfFunctions}

We start by defining equivalences between gates as follows:

\begin{definition}\label{d:GeneralizedTruthValues}
Given two quantum gates $U$ and $V$, we say that

\begin{itemize}
\item $U$ is equivalent to $V$ with respect to
$\rho\in\mathcal{C}(\mathcal{H})$ and $P\in\mathcal{P}(\mathcal{H})$
(and we denote it by $U\equiv^{\rho}_{P}V$) if and only if
$$Tr(U\rho U^{\dagger}P)=Tr(V\rho V^{\dagger}P)$$
For a given quantum gate $U$, we call \emph{probabilistic truth
value associated with $U$ with respect to
$P\in\mathcal{P}(\mathcal{H})$} the real number $Tr(U\rho
U^{\dagger}P)$.
\item
$U$ is equivalent to $V$ with respect to
$\rho\in\mathcal{C}(\mathcal{H})$ (and we denote it by
$U\equiv^{\rho}V$) if and only if
$$Tr(U\rho U^{\dagger}P)=Tr(V\rho V^{\dagger}P)$$ for all
$P\in\mathcal{P}(\mathcal{H})$.
\item $U$ is equivalent to $V$ with respect to $P\in\mathcal{P}(\mathcal{H})$ (and we denote it by
$U\equiv_{P}V$) if and only if
$$Tr(U\rho U^{\dagger}P)=Tr(V\rho V^{\dagger}P)$$
for all $\rho\in\mathcal{C}(\mathcal{H})$.
\item $U$ is equivalent to $V$ (and we denote it by $U\equiv V$) if and
only if $$Tr(U\rho U^{\dagger}P)=Tr(V\rho V^{\dagger}P)$$ for all
$\rho\in\mathcal{C}(\mathcal{H})$ and
$P\in\mathcal{P}(\mathcal{H})$.
\end{itemize}
\end{definition}

\noindent Notice that neither $U\equiv^{\rho}V$ nor
$U\equiv^{\rho}_{P}V$ imply that $U=V$. On the contrary, $U\equiv V$
implies $U=V$. We have included this last (trivial) definition just
to easily illustrate the fact that $U\equiv^{\rho}V$ and
$U\equiv^{\rho}_{P}V$ are relaxations of the identity relationship.
Indeed, we can summarize this as follows:
$$U=V\Longleftrightarrow U\equiv V\Longrightarrow U\equiv^{\rho}V\Longrightarrow U\equiv^{\rho}_{P}V.$$

\noindent It is easy to find counterexamples showing that the
converse implications are not valid.

The above definitions of equivalence between logical gates have
associated the following probabilistic truth values:

\begin{definition}\label{d:ProbabilisticTruthValues}
Given two gates $U$ and $V$, we say that
\begin{itemize}
\item For a given quantum gate $U$, we call \emph{probabilistic truth value
associated with $U$ in the context $P$ and state $\rho$} the real
number $Tr(U\rho U^{\dagger}P)$.
\item
For a given quantum gate $U$, we call \emph{probabilistic truth
values associated with $U$ in the context $P$ the family of real
numbers $Tr(U\rho U^{\dagger}P)$, where $\rho\in\mathcal{C}$}.
\item For a given quantum gate $U$, we call \emph{probabilistic truth
values associated with $U$ in the state $\rho$ the family of real
numbers $Tr(U\rho U^{\dagger}P)$, where
$P\in\mathcal{P}(\mathcal{H})$}.
\end{itemize}
\end{definition}

\noindent Notice that the above definitions have associated a notion
of implication between gates:

\begin{definition}\label{d:Implications}
Given two gates $U$ and $V$, we say that
\begin{itemize}
\item $U\leq V$ with respect to $\rho\in\mathcal{C}(\mathcal{H})$
and $P\in\mathcal{P}(\mathcal{H})$ (and we denote it by
$U\leq^{\rho}_{P}V$) if and only if
$$Tr(U\rho U^{\dagger}P)\leq Tr(V\rho V^{\dagger}P)$$
\item
$U\leq V$ with respect to $\rho$ (and we denote it by
$U\leq^{\rho}V$) if and only if
$$Tr(U\rho U^{\dagger}P)\leq Tr(V\rho V^{\dagger}P)$$ for all
$P\in\mathcal{P}(\mathcal{H})$.
\item
$U\leq V$ with respect to $P$ (and we denote it by $U\leq_{P}V$) if
and only if
$$Tr(U\rho U^{\dagger}P)\leq Tr(V\rho V^{\dagger}P)$$
for all $\rho\in\mathcal{C}(\mathcal{H})$.
\end{itemize}
\end{definition}

It is also important to notice that the ``$\equiv^{\rho}_{P}$"
relationship is coincident with the notion of probabilistic truth
values of previous publications. For example, in
\cite{DallaChiara-IJQI,ManyValuedApproach}, the probabilistic truth
value of the Toffoli gate is defined as:

\begin{equation}
p(\rho\otimes\sigma)=Tr(T\rho\otimes\sigma\otimes|0\rangle\langle
0|T)(I\otimes I\otimes P_{1}))
\end{equation}

\noindent where $P_{1}:=|1\rangle\langle 1|$.

Another important issue is at stake here. The notion of
probabilistic truth value outlined in Definition
\ref{d:GeneralizedTruthValues} contains the Boolean truth notion as
a particular case as follows. We use as starting state the elements
of the computational basis. Then, implement matrix versions of the
usual classical gates (as for example, Toffoli). Use as a projection
operator the projection associated with any element of the
computational basis (or more generally, one may use the Boolean
lattice $\mathcal{P}_{\mathbf{B}_{0}}$, defined in Section
\ref{s:QuantumComputers}). This procedure yields a Boolean calculus
which is isomorphic to the usual one in reversible classical
computation.

With the above definitions, we can define the truth value associated
to each measurement outcome (i.e., the reading process) of any
quantum protocol. The final truth value of the protocol, associated
to the probability of occurrence of the success subspace, will be
related to the probability of success of the algorithm.

As in the classical case, we need to test the equivalence between
different sets of gates. This can be done in a natural way by
appealing to the $\equiv^{\rho}$ and $\equiv^{\rho}_{P}$ relations.
Indeed, if our aim is to compare the action of two sets of gates
regarding a definite subspace of success, we must use the
$\equiv^{\rho}_{P}$ relationship. If our aim is to compare two gates
regarding the unitary process before any reading
(measurement), we must use the stronger relation $\equiv^{\rho}$.

Notice that this last definition of equivalence (the one given by
$\equiv^{\rho}$) is the generalization of the truth table to the
Boolean setting. Let us elaborate a little bit on this notion. In a
classical setting, if we aim to compare between logical circuits
(defined as compositions of Boolean functions), all we must do is
to list truth tables on each side of the equation. A similar remark
follows for non-deterministic classical computation: we must test
the equivalence by evaluating the probability of each output on each
side of the equation. In both cases (deterministic and
no-deterministic), if these numbers are coincident, we speak about
logically equivalent algorithms.

However, now, as remarked above, we have infinitely many contexts in the
quantum case (indeed, the~whole $\mathcal{P}(\mathcal{H})$ is
available). This forces us to test the equivalence (equality in a
logical circuit equation) in several different contexts. Let us
illustrate this statement  with a concrete example. Suppose that we
have two gates $U$ and $V$, and that we want to check their
equivalence with respect to a reference state $\rho$. Thus, we must
compare $Tr(U\rho U^{\dagger}P)=Tr(V\rho V^{\dagger}P)$ for a
suitably chosen set of projection operators (in principle, there are
infinitely many of them, but in practice, only a reduced family of
them is needed, depending on the dimensionality of the Hilbert
space. This is a well-studied question). Notice that $Tr(U\rho
U^{\dagger}P)=Tr(V\rho V^{\dagger}P)$ is equivalent to $Tr(\rho
U^{\dagger}PU)=Tr(\rho V^{\dagger}PV)$. To test this
equation (and to give the problem  a form similar to the classical
one) we can choose a suitable family of orthonormal bases, each
defining a different measurement context. Each one of these bases
defines its own ``computational basis''. Thus, instead of checking
the probabilities in one single Boolean context (a set of outputs)
as in the non-deterministic classical case, what we are doing here
is to perform the same procedure for all possible quantal contexts
defined by the chosen set of bases (or at least, for~all possible
contexts of interest for the problem we want to solve). All these
properties reveal that quantum computing is the non-commutative
version of classical non-deterministic computation (as expected!),
and that the relationship defined by $\equiv^{\rho}$ is nothing but
the generalization of the truth tables to the non-commutative
setting.

The above definitions introduce different equivalence notions of
gates. In addition, we can compute a kind of quotient of $P(G)$ with regards
to this equivalence relationships (i.e., we can compute the quotient
spaces $P(G)/\equiv^{\rho}_{P}$ and $P(G)/\equiv^{\rho}$). Notice
that $P(G)/\equiv$ is meaningless, because $P(G)/\equiv$ equals
$P(G)$. Each class of $P(G)/\equiv^{\rho}_{P}$ and
$P(G)/\equiv^{\rho}$ contains equivalent gates for the different
purposes defined by the different equivalence relations.

\section{Examples of Quantum Algorithms}\label{s:QuantumAlgorithms}

Now we show how our approach can accommodate some of the most
important quantum algorithms.

\subsection{Deutsch-Jozsa Algorithm}

Let us examine first the Deutsch-Jozsa algorithm
\cite{NielsenandChuang}. In this case, the task is to determine if a
function $f$ is constant or balanced. There are four functions from
$\{0,1\}\longrightarrow\{0,1\}$, namely:
\begin{equation}
\begin{array}{c}
f_{1}(0)=0\,\,\,f_{1}(1)=1\\
f_{2}(0)=1\,\,\,f_{2}(1)=0\\
f_{3}(0)=0\,\,\,f_{3}(1)=0\\
f_{4}(0)=1\,\,\,f_{4}(1)=1\\
\end{array}
\end{equation}

Thus, we have two classes: $C=\{f_{1},f_{2}\}$ and
$B=\{f_{3},f_{4}\}$, and we must determine if the unknown function
$f$ belongs to $B$ or to $C$. Let us see now how this can be reduced
to the steps outlined in Section \ref{s:QuantumComputers}.

\vspace{6pt}
\noindent {Step 1}. In the first step, prepare the quantum
state $|0\rangle|1\rangle$.

\noindent {Step 2}. Next, the Hadamard operator is applied to
both qubits yielding the state:

$$\frac{1}{2}(|0\rangle+|1\rangle)(|0\rangle-|1\rangle).$$

\noindent The quantum implementation of the function $f$ (which
establishes
 the connection between the classical problem and the quantum
computation) will be given by a quantum operator such that it maps
$|x\rangle|y\rangle$ to $|x\rangle|f(x)\oplus y\rangle$. Applying
this function to the state gives

$$(-1)^{f(0)}\frac{1}{2}(|0\rangle+(-1)^{f(0)\oplus f(1)}|1\rangle)(|0\rangle-|1\rangle).$$

\noindent Now, applying the Hadamard transformation again to the
first qubit we get:

$$|\psi\rangle=(-1)^{f(0)}\frac{1}{2}((1+(-1)^{f(0)\oplus f(1)})|0\rangle +(1-(-1)^{f(0)\oplus f(1)})         |1\rangle)(|0\rangle-|1\rangle).$$

\noindent {Step 3}. The next step consists in determining the
projection of the above state to the subspaces represented by
projection operators $|0\rangle\langle 0|\otimes\textbf{1}$ and
$|1\rangle\langle 1|\otimes\textbf{1}$.

\subsection{ Determination of a Function's  Period}

The determination of the period of a periodic function $f$ lies at
the heart of the Shor and Simon quantum computation algorithms
\cite{NielsenandChuang}. The objective now is to determine the
period of a function $f:\mathcal{Z}_{N}\longrightarrow \mathcal{Z}$,
such that $f(x+r)=f(x)$ for all $x$. It is assumed that the function
does not take the same value twice in the same period.

\vspace{6pt}
\noindent {Step 1.} Start the computer by generating -using
the usual procedure- the state:

\begin{equation}
|f\rangle=\frac{1}{\sqrt{N}}\sum_{x=0}^{N-1}|x\rangle|f(x)\rangle.
\end{equation}

\noindent It is not possible to extract the period yet. Even if we
measure the value of the second register and obtain the value
$y_{0}$, we will end up with the following state in the first
register (with $x_{0}$ the smallest $x$ such that $f(x)=y_{0}$ and
$N=K r$):

\begin{equation}
|\psi\rangle=\frac{1}{\sqrt{K}}\sum^{K-1}_{k=0}|x_{0}+k r\rangle .
\end{equation}

\noindent However, $|\psi\rangle$ does not give us information about $r$
yet.

\vspace{6pt}
\noindent {Step 2.} To obtain the period, it is necessary to
apply the quantum Fourier transform (QFT), which is a unitary matrix
with entries

\begin{equation}
\mathcal{F}_{ab}=\frac{1}{\sqrt{N}}\exp^{2\pi i \frac{ab}{N}}.
\end{equation}

\noindent By applying the QFT to $|\psi\rangle$ we obtain

\begin{equation}
\mathcal{F}|\psi\rangle=\frac{1}{\sqrt{r}}\sum^{r-1}_{j=0}\exp^{2\pi
i \frac{x_{0}j}{r}}|j\frac{N}{r}\rangle .
\end{equation}

\vspace{6pt}
\noindent {Step 3.} Finally, a measurement is performed in the
basis $\{|j\frac{N}{r}\rangle\}$, and using the result it is
possible to determine the period of the function as follows. The
obtained value $c$ will be such that $c=j\frac{N}{r}$, for some
$0\leq j\leq r-1$. Then, $\frac{c}{j}=\frac{N}{r}$, and if $j$ is
coprime with $r$, it will be possible to determine $r$. The success
of the algorithm depends on the fact that $j$ and $r$ will be
coprimes with a large enough probability.

\section{Axiomatization of the Quantum Computational Logic}\label{s:Axiomatization}

In the classical case, we know that the functions are given by a
commutative calculus generated by the elementary Boolean functions
$\vee$, $\wedge$ and $\neg$. Thus, the axiomatization of a classical
computational logic can be given in terms of the axioms defining a
Boolean algebra. Is it possible to proceed in an analogous way for
our notion of quantum computational logics? We outline an answer to
this question below.

Notice first that the classical connectives $\vee$, $\wedge$ and
$\neg$ have the natural interpretation given by:

\begin{itemize}
\item $\vee$ is the operation of disjunction in our natural
language.
\item $\wedge$ is the operation of conjunction in our
natural language.
\item $\neg$ is the operation of negation in our
natural language.
\end{itemize}

\noindent However, quantum computers operate in a very different way. If
we consider as elementary gates the set $G_{3}$, we have the
following semantical/physical interpretation:

\begin{itemize}
\item $H_{i}$: generates a superposition in qubit $i$.
\item $CNOT$: flips the value of the second qubit if the control qubit is
$1$; do nothing otherwise.
\item $R_{\phi}$ adds a phase of $\phi$ in one of the terms of the superposition.
\end{itemize}

Notice the following: all truth values of a classical probabilistic
calculus are computed regarding classical propositions. However,
these propositions are nothing but the elements of a Boolean
algebra. As an example, consider a computer with $N$ bits. Thus, all
possible inputs are given by elements of the set $\{0,1\}^{N}$ (and
also the outputs). Thus, all possible readings (measurements) in the
output of a computation are given by elements of $\{0,1\}^{N}$, and
all possible propositions are nothing but conjunctions,
disjunctions, and negations of this outcome set. In other words, all
elementary propositions are given by the elements of the Boolean
algebra $\mathcal{P}(\{0,1\}^{N})$. The nature of the classical
(deterministic) computing process assigns a valuation to the set
$\{0,1\}$ to each element of the set $\mathcal{P}(\{0,1\}^{N})$.
Non-deterministic classical computation instead, assigns probability
values in the interval $[0,1]$.

The quantum setting, despite of its formidable appearance, is
completely analogous. We have stressed above that the quantum
calculus contains the classical one as a special case. This is
totally reasonable: quantum mechanics defines a probabilistic
classical theory for each context. In other words, a quantum state
can be seen as a collection of classical probability distributions
pasted in a harmonic way using the density operator associated with
that state \cite{Holik-AOP-2014}. What is thus the analogous of a
reading of a quantum register? The answer was given by von Neumann
in the first axiomatization of the formalism: each elementary
reading (i.e., any possible empirically testable proposition) is
represented by a projection operator $P\in\mathcal{P}(\mathcal{H})$.
Thus, we have a very clear operational interpretation of the
equivalence relations and truth values defined in Definition
\ref{d:GeneralizedTruthValues}: what we are doing here is to compute
the truth value of a quantum state regarding a particular
proposition. This notion of truth in quantum computation is all that
we need to compare different sets of gates and the success
probability of each quantum algorithm.

Another important thing to remark is that the calculus defined by
the matrices involved in a quantum case is not commutative. Thus, we
can anticipate that the axiomatization will not involve a Boolean
algebra.

Thus, to define a suitable algebraic axiomatics for quantum
computation, all we must do is to consider: (i) a set $\mathcal{L}$
of empirically testable propositions (which are intended to
represent all possible readings of the quantum register). It is
natural to demand that $\mathcal{L}$ should be  an orthomodular
lattice (as we show below, this covers the classical and quantum
cases as well); (ii) a set of states $\mathcal{C}$ assigning
definite probabilities to each element of $\mathcal{L}$ (these can
be defined in the usual way as $\sigma$-additive
probabilities)\,\,\cite{Holik-AOP-2014}; (iii) an elementary set $G$
of operations acting by automorphisms in $\mathcal{L}$. Remember
that an automorphism of a logic is defined as a bijective map
$U:\mathcal{L}\longrightarrow\mathcal{L}$ satisfying (a)
$U(\mathbf{0})=\mathbf{0}$ and $U(\mathbf{1})=\mathbf{1}$, (b) for
any denumerable sequence $X_{1},X_{2},\ldots$ we have
$U(\bigvee_{n}X_{n})=\bigvee_{n}U(X_{n})$ and
$U(\bigwedge_{n}X_{n})=\bigwedge_{n}U(X_{n})$ and (c) for all
$X\in\mathcal{L}$, $U(X^{\bot})=(U(X))^{\bot}$. The set $G$
generates the collection $P(G)$ of logical polynomials (by
composition). The set $\Xi=\langle\mathcal{L};\mathcal{C};G\rangle$
will be called a \emph{generalized computational scheme}. In the
rest of this section, we make extensive use of the following
notation: given an automorphism $U$ acting on $\mathcal{L}$, define
the action $U(\nu)$ of $U$ on the state $\nu$ as
$U(\nu)(X):=\nu(U(X))$, for all $X\in\mathcal{L}$. Using $\Xi$ we
can define $P(G)$ (that determines the set of all possible logical
gates) and the notions of probabilistic truth value relative to a
context and equivalence of logical gates as follows:

\begin{definition}\label{d:GeneralizedTruthValues}
Given two gates $U,V\in P(G)$, we say that

\begin{itemize}
\item $U$ is equivalent to $V$ with respect to $\nu\in\mathcal{C}$
and $X\in\mathcal{L}$ (and we denote it by $U\equiv^{\nu}_{X}V$) if
and only if
$$\mu(X)=\mu'(X)$$ where $\mu(-)=U(\nu)(-)$ and $\mu'(-)=V(\nu)(-)$.
\item
$U$ is equivalent to $V$ with respect to $\nu$ (and we denote it by
$U\equiv^{\nu}V$) if and only if
$$\mu(X)=\mu'(X)$$ for all $X\in\mathcal{L}$.
\item
$U$ is equivalent to $V$ (and we denote it by $U\equiv V$) if and
only if $\mu(X)=\mu'(X)$ for all $\nu\in\mathcal{C}$ and
$X\in\mathcal{L}$.
\end{itemize}
\end{definition}

\noindent We can also define the notion of \emph{generalized
protocol} using the following steps:

\begin{itemize}
\item {Step 1}. Chose an initial reference state
$\nu\in\mathcal{C}$ (this state is intended to be the same for all
possible algorithms, and is interpreted as the initial state of the
devise).
\item{Step 2}. Apply a collection of gates $\{U_{i}\}_{i=1,...,n}$ to reach a desired final state
$\mu(-)=(U_{n}\cdots U_{2}U_{1})(\nu)(-)$, possessing the properties
needed to perform the desired computation (answer the
question that we need to answer).
\item {Step 3}. Perform a measurement on the system when the state $\mu$ is reached, check the result obtained,
and depending on the result, stop the process, or continue the
protocol if necessary (which will involve a similar -in some cases,
the same- process).
\end{itemize}

\noindent The intended interpretations of the above notions are
similar to those of classical and quantum algorithms. We recover
classical computation by setting $\mathcal{L}=\mathcal{B}$ (where
$\mathcal{B}$ is a Boolean algebra) and quantum computation when
$\mathcal{L}=\mathcal{P}(\mathcal{H})$ (using the concomitant
definitions of initial reference state, probabilities, etc.).

As in the quantum case, we can give definitions of
probabilistic truth values:

\begin{definition}\label{d:ProbabilisticTruthValues}
Given two gates $U$ and $V$, we say that
\begin{itemize}
\item For a given generalized gate $U$, we call \emph{probabilistic truth value
associated with $U$ in the event $X\in\mathcal{L}$ and initial state
$\nu$} the real number $U(\nu)(X)$.
\item
For a given generalized gate $U$, we call \emph{probabilistic truth
values associated with $U$ in the event $X\in\mathcal{L}$} the family
of real numbers $U(\nu)(X)$, where $\nu\in\mathcal{C}$.
\item For a given generalized gate $U$, we call \emph{probabilistic truth
values associated with $U$ in the state $\nu$} the family of real
numbers $U(\nu)(X)$, where $X\in\mathcal{L}$.
\end{itemize}
\end{definition}

\noindent Notice that the above definitions have associated a notion
of implication between gates:

\begin{definition}\label{d:Implications}
Given two gates $U$ and $V$, we say that
\begin{itemize}
\item $U\leq V$ with respect to $\nu\in\mathcal{C}$
and $X\in\mathcal{L}$ (and we denote it by $U\leq^{\nu}_{X}V$) if
and only if
$$U(\nu)(X)\leq V(\nu)(X)$$
\item
$U\leq V$ with respect to $\nu$ (and we denote it by $U\leq^{\nu}V$)
if and only if
$$U(\nu)(X)\leq V(\nu)(X)$$ for all $X\in\mathcal{L}$.
\item
$U\leq V$ with respect to $X\in\mathcal{L}$ (and w denote it by
$U\leq_{X}V$) if and only if
$$U(\nu)(X)\leq V(\nu)(X)$$ for all $\nu\in\mathcal{C}$.
\end{itemize}
\end{definition}


The effect of the Hadamard gate in the computational basis is to
generate a superposition out of the input qubit. Thus, it is
relevant for our construction to describe how superpositions are
defined in arbitrary orthomodular lattices (we follow
\cite{vadar68}, Chapter III, Section 4). Given an orthomodular
lattice $\mathcal{L}$, let $\mathcal{D}$ be a collection of states
in $\mathcal{L}$. Then, a state $\nu$ is a \emph{superposition} of
the states in $\mathcal{D}$, if and only if, for all
$X\in\mathcal{L}$ we have

\begin{equation}
\forall\,\,\mu\in\mathcal{D}\,\,(\mu(X)=0)\,\,\Longrightarrow\,\,\nu(X)=0
\end{equation}

\noindent The above definition coincides with the usual one for the
case of the lattice of projection operators acting on a Hilbert
space and pure states. In Boolean algebras, no pure state can be a
superposition of other pure states. As automorphisms of a logic are
angle-preserving (i.e., they are straightforward generalizations of
the rotations acting on the projective geometry associated with a
Hilbert space), their~effect on a given set of propositions will be,
in general, to generate superpositions. This~fact guarantees that we
will be able to recover a quantum-like computation rich enough so as
to display contextuality and entanglement (as is the case for
example, with factor von Neumann algebras, for~which a version of
the Kochen-Speker theorem can be proved \cite{Doring-KSVNA}; for the
study of correlations in the algebraic approach see
\cite{Halvorson}).

\section{Conclusions}\label{s:Conclusions}

In this work we have presented a generalization of quantum
computational logics capable of dealing with some important examples
of quantum algorithms. We show that our constructions generalize
previous studies on the subject. In Section \ref{s:Axiomatization}
we have outlined an axiomatization for quantum computational logics.
Our developments lead to new problems related to the algebraic
characterization of computational gates in a non-commutative
setting. They also open the door for further generalization of the
notion of \emph{algorithm}, beyond the classical and standard
quantum mechanical formalisms.
\vspace{6pt}

\vskip1truecm

\noindent {\bf Funding} \noindent This work has been partially
supported by Regione Autonoma della Sardegna in the framework of the
project ``Time-logical evolution of correlated microscopic systems''
(CRP 55, L.R. 7/2007, 2015), by Fondazione Banco di Sardegna $\&$
Regione Autonoma della Sardegna in the framework of the project
``Science and its logics, the representation's dilemma'', cup:
F72F16003220002, and by Fondazione Banco di Sardegna of the project
``Strategies and Technologies for Scientific Education and
Dissemination'', cup: F71I17000330002.

\vskip1truecm

\noindent {\bf Acknowledgments} \noindent This work has been
partially supported by CONICET (Argentina).





\end{document}